\documentclass[prl, aps, twocolumn,floats, superscriptaddress, showpacs]{revtex4}
\usepackage{graphicx}
\usepackage{dcolumn}
\usepackage{bm}
\usepackage{color}

\begin{document}

\author{Ping Nang Ma}
\affiliation{Center of
Theoretical and Computational Physics and Department of Physics, The University
of Hong Kong, Hong Kong, China}
\author{Kai Yu Yang}
\affiliation{Institut Theoretische Physik, ETH Z\"urich, 8093 Z\"urich, Switzerland}
\affiliation{Center of
Theoretical and Computational Physics and Department of Physics, The University
of Hong Kong, Hong Kong, China}
\author{Lode Pollet}
\affiliation{Institut Theoretische Physik, ETH Z\"urich, 8093 Z\"urich, Switzerland}
\author{J. V. Porto}
\affiliation{ JQI, National Institute of Standards and Technology, Gaithersburg, MD 20899-8424}
\author{Matthias Troyer}
\affiliation{Institut Theoretische Physik, ETH Z\"urich, 8093 Z\"urich, Switzerland}
\author{Fu Chun Zhang}
\affiliation{Center of Theoretical and Computational Physics, The University of Hong
Kong, China}
\affiliation{Department of Physics, The University of Hong Kong, China}
\title{Influence of the trap shape on the superfluid-Mott transition in
ultracold atomic gases}
\title{Detection of a cleaner superfluid-Mott transition in realistic
trapping potentials}
\title{Influence of the trap shape on the detection of the superfluid-Mott
transition}

\begin{abstract}
The coexistence of superfluid and Mott insulator, due to the quadratic confinement potential in current optical lattice experiments, makes the accurate detection of the superfluid-Mott transition difficult. Studying alternative trapping potentials which are experimentally realizable and have a flatter center, we find that the transition can be better resolved, but at the cost of a more difficult tuning of the
particle filling. When mapping out the phase diagram using local probes and the local density approximation we find that the smoother gradient of the parabolic trap is advantageous.
\end{abstract}

\maketitle

%%%%%%%%%%%%%%%%%%%%%%%%%% AUTHORS %%%%%%%%%%%%%%%%%%%

%%%%%%%%%%%%%%%%%%%%%%%%%%%%%%%%%%%%%%%%%%%%%%%%%%%
%\title{Obtaining a cleaner superfluid-Mott transition with flatter trap shapes}

%A square well trap, even of modest size, would give a homogeneous phase and a very clear signal at the transition but its implementation is not straightforward.

%%%%%%%%%%%%%%%%%%%%%%%%%%%%%%%%%%%%%%%%%%%%%%%%%%%
Ultracold atomic gases in optical lattices provide a tunable and well-controlled realization of strongly correlated quantum lattice models such as the bosonic or fermionic Hubbard models \cite{Review}. Unlike strongly correlated materials, where the microscopic model is often unknown and the situation is complicated by additional interactions such as phonons and long-range Coulomb interactions, cold atoms provide a clean realization of prototypical models of strongly correlated systems.

In a seminal paper, Greiner {\it et al.} \cite{Greiner} demonstrated  the loss of coherence in a Bose-Hubbard model as one tunes the interaction parameters from a superfluid Bose-Einstein condensate (BEC) to a bosonic Mott insulator (MI) in a parabolic trap. While this experiment beautifully demonstrates the BEC and MI phases, it remains at the {\em qualitative} level of demonstrating the existence of the two phases. To use ultracold atomic gases as a quantum simulator for strongly interacting quantum systems one has to go an important step further and {\em quantitatively} determine the phase diagrams.  Apart from devising new detection techniques, there are two major challenges to be solved: thermometry and the influence of the trap shape. Here we focus on the latter aspect. 

While the Greiner {\it et al.} experiment identifies the BEC and MI phases, the ``transition'' is a smooth and broad crossover between these two limits instead of a true phase transition due to the inhomogeneous parabolic trapping potential \cite{Wessel}. The coherence fraction (i.e., the brightest spot of the interference pattern) does not display any distinct features upon emergence of a central Mott plateau, but decreases smoothly over a broader range than expected for the uniform system. Similarly, the FWHM barely changes near the Mott onset, but increases abruptly for much deeper lattices. While for the homogeneous lattice the transition can be more or less accurately determined using the coherence fraction and the FWHM, no such approach is possible in the presence of a parabolic trap, and the Mott onset cannot simply be inferred from time-of-flight experiments~\cite{Wessel}. 

We can attempt to obtain better information about the phase transition of the homogeneous system by studying the crossover in a trap if we use ``flatter'' traps with steeper walls, mimicking a homogenous system with hard walls, since in such a trap the crossover is expected to be sharper and closer to the true phase transition.

In this Letter we investigate the effect of the shape of the trapping potential $V(\vec{r})$ on the detection of the BEC-MI transition in two-dimensional bosonic lattice gases, described by the single-band Bose Hubbard Hamiltonian~\cite{Jaksch98}:
\begin{equation}
\hat{H} = -t \sum_{\langle i,j \rangle}\left( \hat{b}_i^{\dagger} \hat{b}_j
+ \mathrm{h.c.} \right)+ \frac{U}{2} \sum_i \hat{n}_i(\hat{n}_i-1) + \sum_i
V(\vec{r}_i) \hat{n}_i.
\end{equation}
For the hopping amplitude $t$ and the on-site repulsion strength $U$ we use the microscopic parameters in the standard tight-binding approximation \cite{Jaksch98} for ${}^{87}\mathrm{Rb}$ atoms with an $s$-wave scattering length of $a_s = 102(6)a_0$~\cite{Buggle04}, where $a_0$ is the Bohr radius. We obtain a two-dimensional lattice by integrating over the strong harmonic confinement in the $z$-direction with oscillator length $a_z = 53.5 \mathrm{nm}$ and assume a lattice laser with wavelength $\lambda = 820 \mathrm{nm}$, which gives a lattice spacing $a=\lambda/2$. The recoil energy $E_{\mathrm{R}} = h^2/(2m\lambda^2)$ is the natural unit to express laser intensities. 

\begin{figure}
\includegraphics[width=6cm]{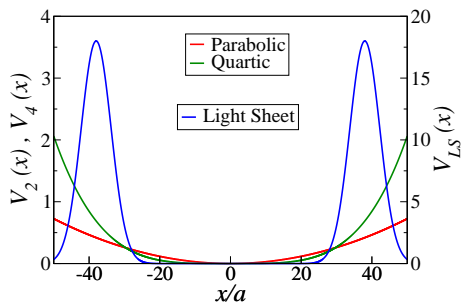}
\caption{(Color online). The three trapping potentials used in our comparison along one direction. The scale is different for each of the potentials, in particular the walls of the light sheet potential are much steeper than suggested in the figure. }
\label{fig:trap}
\end{figure}

We will investigate three different trapping potentials: parabolic $V(\vec{r} ) = V_2 (r/a)^2$, quartic  $V( \vec{r}) = V_4 (r/a)^4$ and flat potentials with ``light sheet'' walls. The quartic potential can be realized either using supergaussian laser beams or by compensating the typically quadratic part of the trapping potential with a blue-detuned laser. An even flatter trap center can be obtained in a ``light sheet'' trap using repulsive walls of blue-detuned Gaussian laser beams. The waist of the light sheet laser beam is set to be 12 lattice spacings, {\it i.e.} $\sigma = 6a$. The trapping potential of these light sheets is
\begin{eqnarray}
V (\vec{r}) &=& V_{LS} [\exp (-\frac{(x - r_c)^2}{\sigma^2}) \,+\, \exp (-%
\frac{(x + r_c)^2}{\sigma^2})  \nonumber \\
& & +\, \exp (-\frac{(y - r_c)^2}{\sigma^2}) \,+\, \exp (-\frac{(y + r_c)^2}{%
\sigma^2}) \,],
\end{eqnarray}
where $\vec{r}=(x,y)$ is the distance from the center of the trap and we have placed light sheets parallel to the $x$ and $y$-axis centered along $x=\pm r_c$ and $y=\pm r_c$. These trapping potentials are schematically shown in Fig.~\ref{fig:trap}.

In all cases we assume that the whole confinement is coming from the external potential only and that there is no contribution from the lattice laser waists  or the $z$-direction confining potential. This can in principle be done by carefully compensating the Gaussian trapping potential of the waist of a red-detuned lattice laser with a blue-detuned dipole trap. The required homogeneity for the lattice potential depends on the energy and length scales of the problem. Globally, the lattice potential should be flat to much better than the local chemical potential at the transition, which for our $^{87}$Rb example is set by $U$ and should be smaller than $ E_{\mathrm{R}}$ over the $~50\times50$ sites in the sample.  Locally, the site-to-site potential should be flat to much better than $t$ at the transition,  thus variations should be much smaller than $ 0.05 E_{\mathrm{R}} $ over the $a=410$~nm lattice spacing. We note that for ``super-exchange" physics which scales as $t^2/U$, these requirements are much more stringent. In any case, using larger gaussian beams provides significantly more homogeneous lattices, at the expense of increased required laser power.

\begin{figure}
\includegraphics[width=\columnwidth]{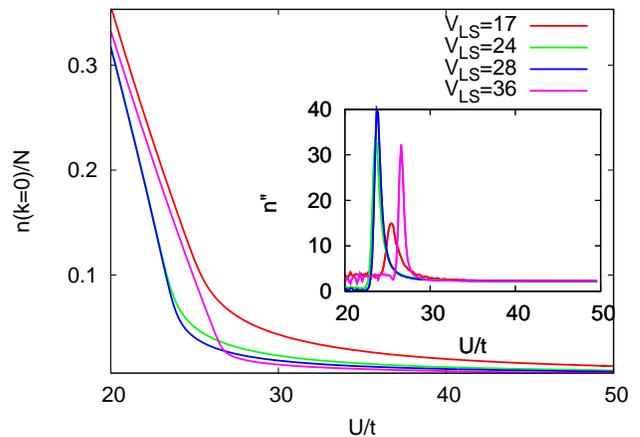}
\caption{(Color online).  Condensate fraction (main figure) and its second derivative (inset)
as a function of $U$ for a number of different fillings and a light sheet
trap with amplitude $V_{LS} = 18 E_{\mathrm{R}}$. The light sheet is
centered at $r_c = (0, \pm 38a)$ and $(\pm 38a, 0)$ and there are 2550
bosons. For comparison we include the condensate fraction in a homogeneous
system. The inset shows the second derivative $n'' = (1/N) d^2n(k=0)/d(U/t)^2$ for the same curves. }
\label{fig:LS_1}
\end{figure}

We investigate the superfluid-Mott transition in these trapping
potentials using Gutzwiller mean-field theory \cite{MF} at zero temperature with a fixed particle number $N$.  We miminize
the ground state energy $E_{G}=\left\langle G|H|G\right\rangle $ of the
mean-field Gutzwiller ansatz $\left\vert G\right\rangle =\prod_{i}\left(
\sum_{n=0}^{\infty }f_{n}^{\left( i\right) }\left\vert n\right\rangle
_{i}\right) $, where $\left\vert
n\right\rangle _{i}$ denotes the Fock state with $n$
particle numbers at site $i$. While Gutzwiller mean-field
theory does not give the exact value of the critical point, a comparison with an accurate phase diagram  \cite{MFphasediagram} shows that 
the mean-field approach captures the
physics of the BEC-MI phase transition and is reliable for the present purpose of
qualitatively comparing the effects of different trapping potentials. 

{\it Estimating the transition from the coherence fraction } -- In a homogeneous system the Mott transition is clearly observed by the vanishing condensate fraction $f_c=n(\vec{k}=0)/N$. 
%In a trapped system the smooth crossover
%due to phase coexistence makes the definition of the transition point arbitrary and the best one can do is consistently apply a fixed procedure to define a crossover point as estimate for the true transition.
%Guided by the homogeneous system which has a cusp in $n(\vec{k}=0)$ at the
%transition we determine the ``transition'' in the trapped system as the
%point where the second derivative of the condensate fraction reaches a
%maximum \cite{Wessel}. As we see in Fig.~\ref{fig:LS_1} this procedure gives numbers in good
%agreement with the transition point in a homogeneous lattice. 
In a trapped system the smooth crossover due to phase coexistence makes the definition of the transition point somewhat ambiguous. To estimate it, we note that  there is a cusp in $n(\vec{k}=0)$ at the transition point in the homogeneous system. We determine the "transition" point in the trapped system as the point where the second derivative of the condensate fraction reaches a maximum \cite{Wessel}. As we can see form the inset in Fig.~\ref{fig:LS_1} this gives a 'transition point' in good agreement with that in a homogeneous lattice.

\begin{figure}
\includegraphics[width=\columnwidth]{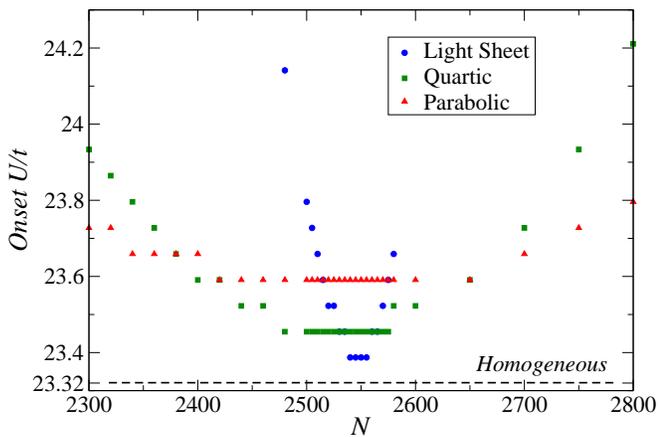}
\caption{(Color online).  The estimate for Mott transition as a function of the total number
of particles $N$ for parabolic, quartic and light sheet
traps. The values for the trapping potentials are chosen as $V_2 = 2.9
\times 10^{-4} E_{\mathrm{R}}, V_4 = 3.3 \times 10^{-7} E_{\mathrm{R}}$, and
$V_{\mathrm{LS}} = 18 E_{\mathrm{R}}$ in order to have Mott domains of
comparable size and the minimum to occur at similar particle numbers. The
dashed line indicates the critical point as obtained in the same mean-field
approximation. }
\label{fig:sensitivity}
\end{figure}

\begin{figure}[b]
\includegraphics[width=\columnwidth]{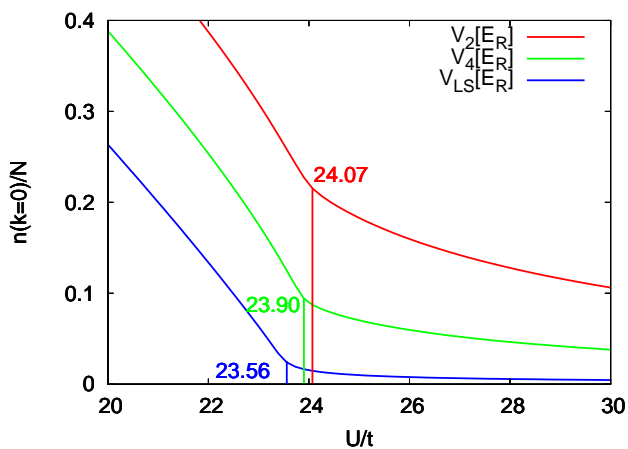}
\caption{(Color online).  Comparison between the optimal curves for the parabolic, quartic
and light sheet potential. It is very hard to determine the transition point with the parabolic trap. The light sheet trap gives the best estimate for the phase transition.}
\label{fig:comparison}
\end{figure}

We find a strong dependence of the critical point on the particle number, as shown in Fig. \ref{fig:sensitivity}, which is important in view of the difficulty of controlling particle number in experiments. Using the minimum of the transition $U/t$ as a function of particle numbers as an estimate for the tip of the Mott lobe, we find only a small dependence on the trap shape. The condensate fraction and the estimate for the transition point at these minima are shown in  Fig.~\ref{fig:comparison}. Changing a parabolic trap to a quartic or light sheet trap decreases the deviation compared to the homogeneous system from 1.2\% to 0.6\% and 0.3\%, respectively.

\begin{figure}
\includegraphics[width=\columnwidth]{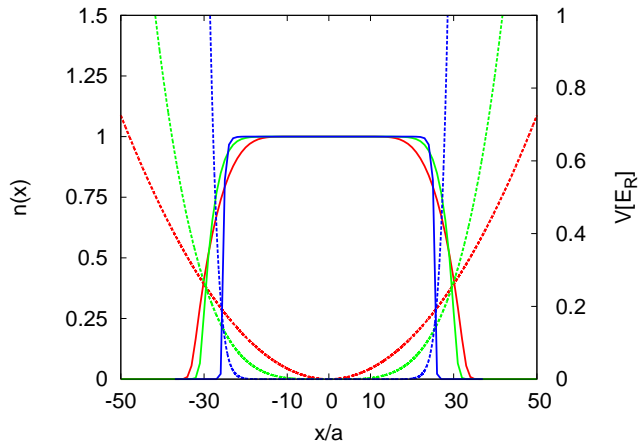}
\caption{ (Color online).  Density profile (left, full line) and trapping potential (right,
dashed line)  for $U/t$ at the tip of the Mott lobe for $N=2550$ particles. }
\label{fig:density}
\end{figure}

Looking at density profiles in Fig.~\ref%
{fig:density} we see that when a central Mott insulating region is formed at the tip of the Mott lobe, the width of the surrounding superfluid region in the light sheet setup is considerably smaller than for the quadratic and quartic trap. The density in the superfluid region also drops much faster as can be expected from the shape of the potential. This is reflected in the condensate fraction curves of Fig.~\ref{fig:LS_1}, where we also compare to the homogeneous case. When $U/t$ is large, the light sheet potential has a much lower condensate fraction than the quartic or parabolic potential,  because the superfluid edges are almost absent. 
%As we have seen the estimate for the transition shifts towards the one of the homogeneous lattice when we make the trap flatter. 
The gradient change in $n(k=0)/N$ with $U$ is more abrupt for the light sheet potential than for more curved potentials. Other quantities than the coherent fraction that are also related to the TOF interference pattern measurements, such as the FWHM of the first peak or the visibility $(n_{\mathrm{max}} - n_{\mathrm{min}})/(n_{\mathrm{max}} + n_{\mathrm{min}})$ are also insensitive probes for the superfluid-Mott crossover in a parabolic trap. 
%The reason is that the presence of a superfluid edge of even modest size gives rise to a big coherence fraction, which may mask the presence of a potentially sizeable Mott domain in TOF experiments. 
We have checked with quantum Monte Carlo that for the light sheet potential, we obtain FWHM and visibility curves that are closer the homogeneous lattice results -- again confirming that a flatter trap is advantageous if atom number and temperature can be well controlled.

The main advantage of the light sheet trap is thus the much sharper signal (compare the bends in the curves of Fig. \ref{fig:comparison}). The light sheet trap is, however, also the most sensitive one to particle number fluctuations. While the transition in the parabolic trap varies by less than 1\% when changing $N$ by 10\%, the transition in the light sheet trap varies as much when changing $N$ by only 1\%. It is clear that quasi-flat traps such as the light sheet trap are extremely sensitive to the particle number; the transition to the Mott phase is largely density-driven; a big value of $U/t$ alone is not sufficient for observing the Mott state in flat traps. Another challenge comes from entropy considerations, which must be taken into account since ramping up the optical lattice is to a good approximation an adiabatic and not an isothermal process. Since for the commensurate, homogeneous lattice the Mott gap opens as $\sim U/t$, temperature has to increase proportionally. For the parabolically trapped Bose-Hubbard model, temperature almost scales with $U/t$, but with a pre-factor that is  typically an order of magnitude lower thanks to the non-zero superfluid volume fraction ~\cite{Lode_S}. In the latter case, temperature effects play only a minimal role, but for the light sheet potential they might be important and require a colder initial state that has a much lower entropy. However, once 1\% accuracy in atom number can be achieved experimentally,  a quartic or light sheet trap will give a substantially easier to observe experimental signature and a better estimate for the critical point.

%Despite the lower superfluid volume fraction it is somewhat surprising that
%the condensate fraction of Fig.~\ref{fig:LS_1} is qualitatively not
%different from the parabolic or quartic trap (see ??? and Fig.~\ref%
%{fig:comparison} ).  {\bf [MT: Lode, this is confusing - the magnitude at the transition is substantially smaller]} As noted before \cite{noted}, we see that the interference
%pattern is only meaningful deep in the superfluid and deep in the Mott
%phase; otherwise the condensate fraction is too big. This has a strong
%influence all other related quantities used for observation, such as the
%visibility $(n_{\mathrm{max}} - n_{\mathrm{min}})/(n_{\mathrm{max}} + n_{%
%\mathrm{min}})$ or the full width at half maximum of the momentum distribution.
% We see in Fig.~\ref{fig:FWHM} that the slopes of the FWHM (not taking the Wannier function into account) as a function of $U/t$ vary the most for the light sheet potential. Thus, not only for the condensate fraction but also for the FWHM gives the light sheet potential the relatively best results for locating the transition point.
%\\

\begin{figure}
\includegraphics[width=0.8 \columnwidth]{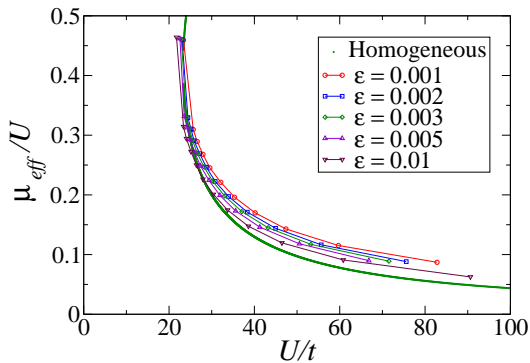}
\caption{ (Color online).  Phase diagram obtained using the local density approximation for
the parabolically trapped two-dimensional Bose-Hubbard model. The phase boundary is determined by looking where the density deviates from unity by a threshold value $\epsilon$. We plot the results obtained for various values of $\epsilon$: $\epsilon=0.001$ to  $\epsilon=0.010$. 
The procedure fails close to the tip of the lobe. }
\label{fig:phase_diagram}
\end{figure}

{\it Obtaining the phase diagram from the local density} -- A new and promising detection technique is the high-resolution quantum gas
microscope {\cite{Greinerxxx} } , giving access to in situ measurements. Using the local chemical
potential approximation, $\mu_{\mathrm{eff}}(i) = \mu - V(i)$, where $\mu$ is the chemical potential, one can map
out the grand canonical phase diagram since for every value of $U/t$ one obtains a slice of
the $( U/t,\mu/U)$ phase diagram. We want to investigate which trapping
potential is best suited for obtaining the phase diagram this way. 

Again the inhomogeneous trap smears out the transition and the local chemical potential
approximation is not valid near the phase boundaries \cite{Wessel}. There is thus a degree of arbitrariness (or uncertainty) in the procedure:
looking at the density profiles of Fig.~\ref{fig:density}, we have to
define where  the Mott insulator ends. 
%One criterion could be the deviation of the density from an integral value. In
%our numerical calculations, it turned out to be better to look at the profile of the
%density variations $\mathrm{Var}(\hat{n_i}) = \langle \hat{n}_i \hat{n}_i
%\rangle - \langle \hat{n}_i \rangle^2$.  $\mathrm{Var}(\hat{n_i})$ is small $(\sim 0)$ inside the Mott insulator and large $(\sim \langle n_i \rangle^2)$  in the superfluid region. Choosing a threshold $\mathrm{Var}(\hat{n_i}) < \epsilon$ as the criterion for detecting the phase transition we plot the obtained phase diagram in  Fig.~\ref{fig:phase_diagram}. 

One criterion could be the deviation of the density from an integral value. Choosing a threshold $\vert \hat{n_i} -1 \vert  < \epsilon$ as the criterion for detecting the phase transition to the $n=1$ Mott lobe, we plot the obtained phase diagram in  Fig.~\ref{fig:phase_diagram}. Plotting the estimates for various values of $\epsilon$ in a parabolic trap in Fig.~\ref{fig:phase_diagram} gives us an idea of the uncertainty of the phase boundary. The resulting phase diagram is in reasonably good agreement with the homogeneous model. Near the tip of the lobe our procedure does not work: we do not obtain any data points there because the density goes initially down along a line of constant chemical potential starting at the critical point.  It is worth noting that had we determined the phase diagram based on density variations the tip of the lobe could be described accurately. 

It turns out that the parabolic trap is best suited for obtaining the phase diagram using local probes since the trapping potential increases the slowest with distance. This gives the smallest errors for the local potential approximation and provides a finely spaced grid of chemical potentials over a wide region. A quartic or light sheet trap results in substantially larger deviations from the homogeneous result, since using local probes the phase diagram is determined from the edge of the Mott region, which here is in the steep boundary zone of the trap and not in the flat bottom.

{\it Conclusions --}
We have studied the two-dimensional Bose-Hubbard model with a
parabolic, quartic and light sheet trapping potential.  We have seen that the
light sheet potential trap gives a much clearer signal for the superfluid-Mott
crossover. 
%However, the momentum distribution for the different trap shapes
%across the cross-over is very similar, showing again that this quantity
%is not well suited for studying the transition. 
It is however much more sensitive to the total number of atoms in the system since
commensurability is important when the inner part of the trap is almost
flat. Nevertheless, our values for $(U/t)$ at the crossover are in good
agreement with mean-field theory for a homogeneous lattice.
% This sensitivity to filling and the high accuracy needed to find the maximum of the second derivative of the condensate fraction make the momentum distribution a less than ideal probe for the transition.

Measuring the local density combined with a local chemical potential approximation allows an easier determination of the phase diagram.  This procedure allows mapping out the full phase diagram in the $(U/t,\mu/U)$-plane and a parabolic trap is preferred over flatter traps since the flatter slope gives access to more values of the chemical potential and improves the accuracy of the local chemical potential approximation. The obtained phase diagram is in good agreement with the homogeneous phase diagram. 

We thus conclude that currently the most promising procedure to quantitatively determine the phase diagram of the Bose-Hubbard model are local measurements in a harmonic trap -- it will still be a challenge to obtain high accuracy on the phase boundary because of errors due to the local density approximation. Quantum Monte Carlo (QMC) simulations of three-dimensional systems with realistic parameters are in progress to estimate the accuracy of this approach. The position of the tip of the Mott lobe could be determined more accurately using a global measurement of $n(\vec{k})$, but the required accuracy in particle filling and in measuring the momentum distribution to find the largest curvature is a challenge for the experiments.

The ALPS libraries \cite{ALPS} were used for parallelization and 
 the simulations were performed on the Hreidar cluster of
ETH Z\"urich. We thank W. Q. Chen, I. B. Spielman, N. V. Prokof'ev, Henning Moritz and  Tilman Esslinger for useful
discussions. We welcome financial support from RGC of HKSAR and the Swiss
National Science Foundation.

\end{document}